# Outside Looking In_

## Approaches to Content Moderation in End-to-End Encrypted Systems

Authors (In Alphabetical Order)

**Seny Kamara**
**Mallory Knodel**
**Emma Llansó**
**Greg Nojeim**
**Lucy Qin**
**Dhanaraj Thakur**
**Caitlin Vogus**

August 2021

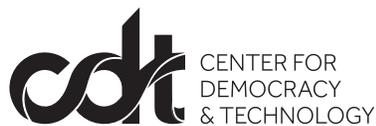

CENTER FOR
DEMOCRACY
& TECHNOLOGY

### SENY KAMARA

Seny Kamara is an Associate Professor of Computer Science at Brown University.

### MALLORY KNODEL

Mallory Knodel is the Chief Technology Officer at the CDT.

### EMMA LLANSÓ

Emma Llansó is Director of the Free Expression Project at CDT.

### GREG NOJEIM

Greg Nojeim is Senior Counsel and Co-Director, Security and Surveillance Project at CDT.

### LUCY QIN

Lucy Qin is a PhD candidate in computer science at Brown University.

### DHANARAJ THAKUR

Dhanaraj Thakur is the Research Director at CDT.

### CAITLIN VOGUS

Caitlin Vogus is Deputy Director of the Free Expression Project at CDT.

# Outside Looking In

## Approaches to Content Moderation in End-to-End Encrypted Systems

**Authors (In Alphabetical Order)**

**Seny Kamara**

**Mallory Knodel**

**Emma Llansó**

**Greg Nojeim**

**Lucy Qin**

**Dhanaraj Thakur**

**Caitlin Vogus**

**WITH CONTRIBUTIONS BY**

Samir Jain, DeVan Hankerson, Hannah Quay-de la Vallee, Air Goldberg, and Tim Hoagland.

**ACKNOWLEDGEMENTS**

We thank Riana Pfefferkorn, Jonathan Lee, and Beda Mohanty for their feedback on an earlier version of this report. We also thank the industry and other experts we talked to who helped inform our analysis. All views in this report are those of CDT.

This work is made possible through a grant from the John S. and James L. Knight Foundation.

**Suggested Citation:** Center for Democracy & Technology (2021). Outside Looking In: Approaches to Content Moderation in End-to-End Encrypted Systems. Center for Democracy & Technology. https://cdt.org/insights/report-outside-looking-in-approaches-to-content-moderation-in-end-to-end-encrypted-systems/

# Contents

# Introduction

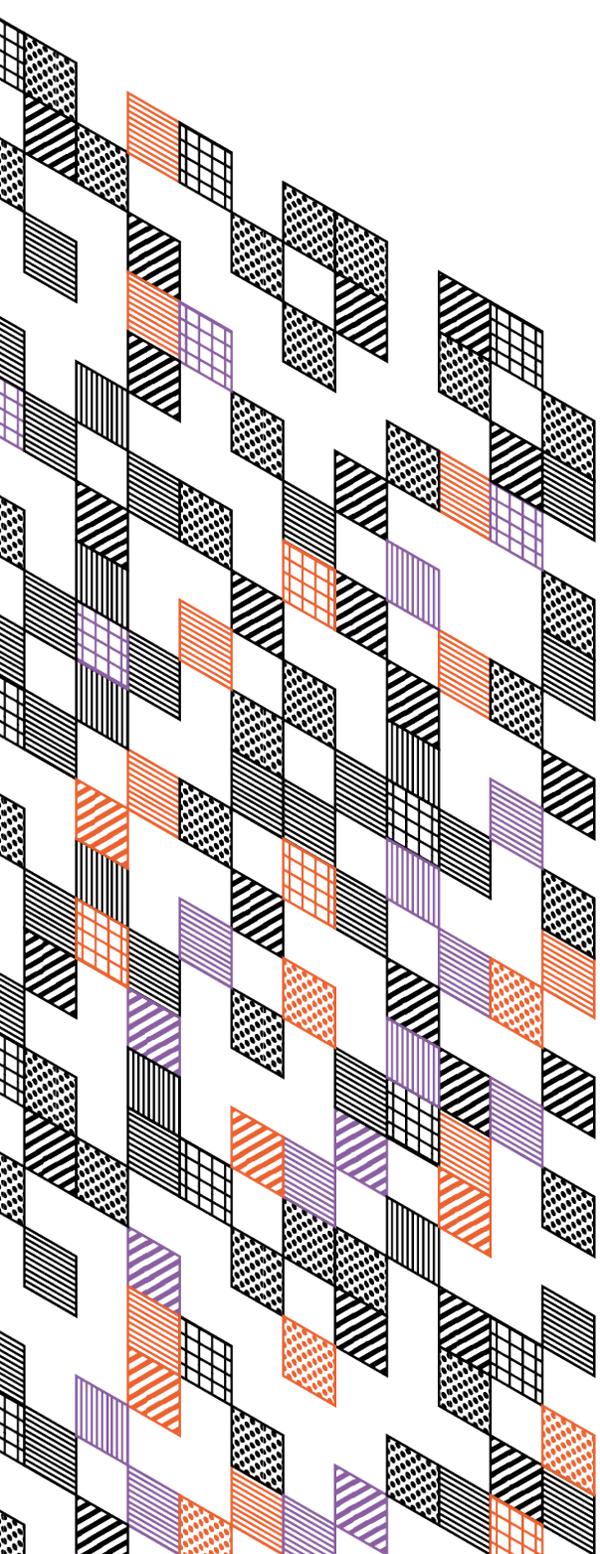

A new front has opened up in the Crypto Wars:[1] *content moderation*. During the 1990's, policy debates in the U.S. and Europe about encryption focused on the benefits and risks of public and foreign access to encryption. Law enforcement and intelligence agencies around the world pushed for restrictions on the development and export of encryption technologies, arguing that greater public access would limit their ability to monitor communications to fight crime and protect the public. In the end, the U.S. government decided against such restrictions with a shift in policy in 1999 (Swire & Ahmad, 2011), and other governments followed suit.

As billions of people around the world began to use encrypted services to protect their privacy and data when communicating with others, the concerns of law enforcement agencies regained prominence in the last decade. In 2014, the then-Director of the FBI argued that encrypted communications were an impediment to law enforcement (Federal Bureau of Investigation, 2014). A 2020 statement by the governments of the U.S., UK, Canada, India, Japan, Australia, and New Zealand expressed similar concerns, calling for greater access by law enforcement to encrypted communications (U.S. Department of Justice, 2020).

Statements such as these tend to focus encryption policy on law enforcement and intelligence agencies' claims that they need to be able to access encrypted communications (National Academies of Sciences, Engineering, and Medicine, 2018). But encryption is not just a law enforcement issue. The availability of secure encrypted communication services is central to privacy, free expression, and the security of today's online commerce (Thompson & Park, 2020).

---

1　See Swire & Ahmad (2011) for a description and history of the policy debates that characterised the Crypto Wars.

Perhaps recognizing the uphill battle they face to undermine such a crucial part of our online infrastructure, some law enforcement officials have begun to link the threat of unconstrained illegal content online to concerns about large social media platforms' content moderation practices. In the U.S., for example, the proposed EARN IT Act was framed as a bill that would establish best practices in content moderation for fighting child sexual abuse material (CSAM), but the debate quickly came to focus on the implications of the bill for end-to-end encryption (E2EE), with many commentators expressing concern that the bill's approach was designed to discourage providers from offering E2EE services or create strong incentives to build in a special access mechanism for law enforcement (Murdock, 2020; Newman, 2020; Ruane, 2020).

But what is the actual effect of encryption on content moderation?

In this paper, we assess existing technical proposals for content moderation in E2EE services. First, we explain the various tools in the content moderation toolbox, how they are used, and the different phases of the moderation cycle, including detection of unwanted content. We then lay out a definition of encryption and E2EE, which includes privacy and security guarantees for end-users, before assessing current technical proposals for the detection of unwanted content in E2EE services against those guarantees.

We find that technical approaches for user-reporting and meta-data analysis are the most likely to preserve privacy and security guarantees for end-users. Both provide effective tools that can detect significant amounts of different types of problematic content on E2EE services, including abusive and harassing messages, spam, mis- and disinformation, and CSAM, although more research is required to improve these tools and better measure their effectiveness. Conversely, we find that other techniques that purport to facilitate content detection in E2EE systems have the effect of undermining key security guarantees of E2EE systems.

The current technical proposals that we reviewed all focus on content detection, which is only one part of the content moderation process. Thus, there may be other useful and effective approaches to moderation for countering abuse in E2EE systems, including user education about applicable policies, improved design to encourage user reports, and consistency of enforcement decisions. These approaches may offer important potential avenues for researchers to build on our analysis.

The legal definition of child sexual abuse material (CSAM) varies by jurisdiction, and generally refers to the depiction or representation of children engaged in sexual activity or abuse (International Centre for Missing & Exploited Children (ICMEC), 2018).

**The availability of secure encrypted communication services is central to privacy, free expression, and the security of today's online commerce.**

# Understanding Content Moderation

Content moderation refers to the set of policies, systems, and tools that intermediaries of user-generated content use to decide what user-generated content or accounts to publish, remove, or otherwise manage (Bloch-Wehba, 2020; see also Grimmelmann, 2015; Klonick, 2018). In this paper, we focus primarily on moderation decisions by hosting intermediaries, though we note the long-running pressure on search engines, message providers, domain name providers, access providers, and other technical intermediaries to engage in moderation.

Content hosts may moderate both content that it is illegal and content that, while legal, violates their terms of service or other rules. Liability frameworks often distinguish between the systems a host has in place to respond to illegal content and those in place to address content that violates their own terms of service. However, in practice, hosts remove substantial amounts of allegedly illegal content as violations of their terms of service (Klonick, 2018). This paper examines the processes hosts may use to take action against user-generated content or user accounts, regardless of the reason.

Hosts take a variety of approaches to content moderation. Some use automated systems to screen user-generated content at upload, for example to detect potential copyright infringement or child sexual abuse material before it is published; others primarily review and moderate content after it has been posted. Some act reactively, reviewing and moderating content only after it is reported as objectionable; others proactively seek out content for moderation (Klonick, 2018). Some rely on manual review by humans to moderate content, while others rely on automated processes.[2] Hosts may use a combination of human and automated review in both reactive and proactive ways (Bloch-Wehba, 2020). But many online services, especially smaller services, continue to rely on reactive, post-publication review of content that is reported to the service operator by a user or other third party.

The distinction between content that is illegal and content that violates a host's terms of service is important; legal regimes requiring the takedown of illegal content should ensure that courts, not intermediaries, are responsible for making the determination that content is illegal before it must be removed. *See* Manila Principles on Intermediary Liability (last visited Mar. 30, 2021), **https://manilaprinciples.org/.**

---

2    For an in-depth examination of the various techniques that may be used to analyze user-generated content, see (Shenkman et al., 2021).

In addition, different hosts exercise different levels of control over content moderation due to the specifics of their site design, business model, ability to incorporate local context into their evaluations, and other considerations (Caplan, 2018). Much attention has been paid to hosts—such as Facebook, Twitter, and YouTube—who are directly involved in content moderation and make moderation decisions centrally. These hosts may write external and internal policies and rules regarding the content that is permitted on their sites, use employees or outsourced contractors to review content and make content moderation decisions, and employ teams to rule on user appeals of content moderation decisions (Gillespie, 2012; Klonick, 2018).

However, other hosts rely on community or distributed moderation in which users themselves moderate content with little or no involvement by the host. For example, Reddit, Wikipedia, Slashdot, and Discord all set baseline policies for content, while relying on volunteers to set additional rules, make content moderation decisions, or both (Caplan, 2018; Grimmelmann, 2015; Lampe & Resnick, 2004; Swartz, 2006). Some hosts combine both methods, employing central decision making for certain content moderation decisions and community moderation for others. For example, the video-streaming service Twitch describes its moderation approach as "a layered approach to safety - one that combines the efforts of both Twitch (through tooling and staffing) and members of the community, working together" (Twitch, 2021).

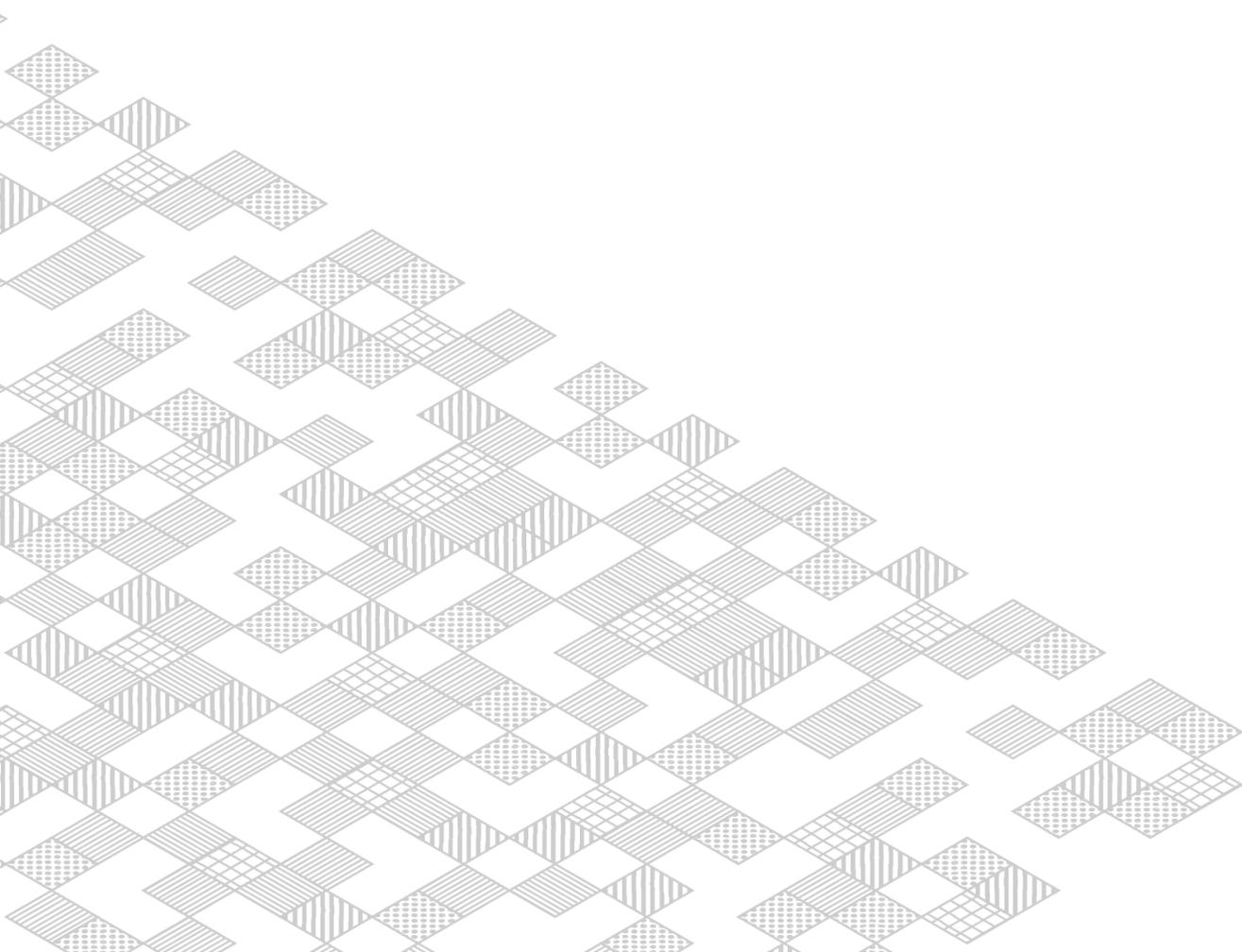

# Phases of Content Moderation

**Content moderation is a much more complex process than simply making binary decisions to either take down or allow user-generated content on a service.**

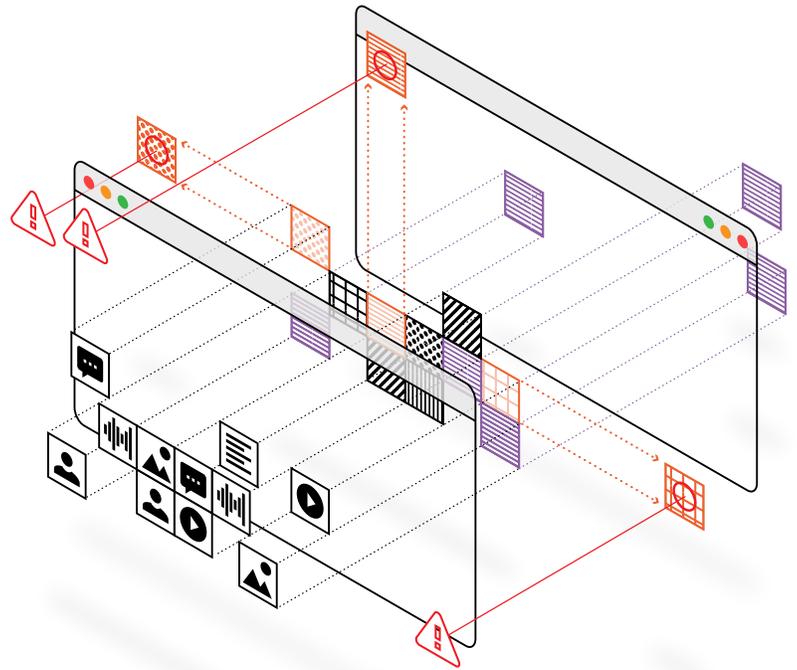

Content moderation is a much more complex process than simply making binary decisions to either take down or allow user-generated content on a service. It is useful to think of content moderation as occurring in six phases: definition, detection, evaluation, enforcement, appeal, and education. Moreover, content moderation is an iterative process; these phases are interrelated, and each phase may happen multiple times and in a different order than described below.

In the **definition** phase, hosts or others determine what user-generated content is and is not permitted on the service. This involves both defining impermissible content and behavior and describing such content and behavior to others, both externally to users and internally. Hosts may define and communicate permissible and impermissible content in their terms of service or community guidelines, but rules may also be defined and communicated in other ways. For example, subreddit rules on Reddit are commonly displayed to users on the subreddit itself, with short phrases to identify the rule topic and a brief explanation (Fiesler et al., 2018).

## A. Definition

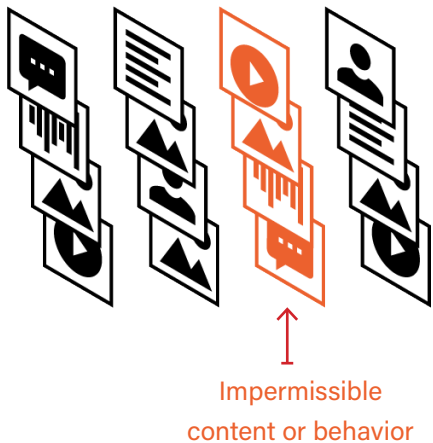

↑
Impermissible
content or behavior

## B. Detection

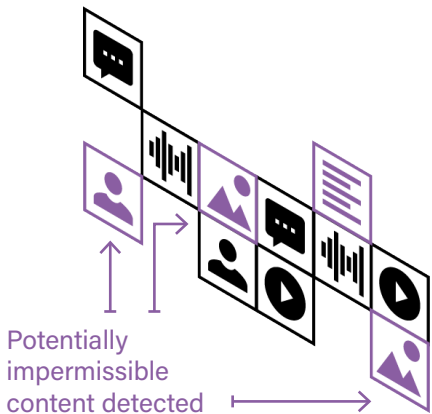

Potentially impermissible content detected

## C. Evaluation

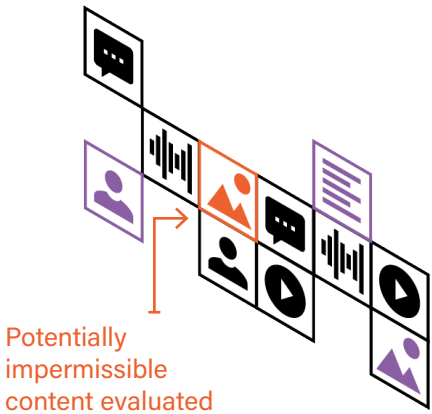

Potentially impermissible content evaluated

## D. Enforcement

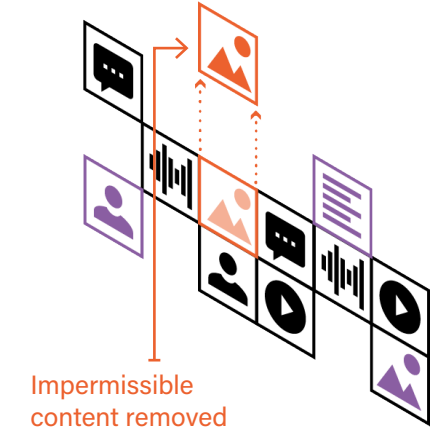

Impermissible content removed

**Detection** is how hosts or other moderators identify user-generated content that may violate their policies or the law (for more on detection, see "Detecting Content in E2EE Environments," p. 15). Hosts engage in a variety of methods to detect content for moderation and may use multiple methods simultaneously. As described above, detection can take place at different points in time—either "before content is actually published on the site, as with ex ante moderation, or after content is published, as with ex post moderation" (Klonick, 2018, p. 1635). Ex post detection may be reactive, in which moderators rely on users or other third parties to "bring the content to their attention," such as through flagging, or proactive, "in which teams of moderators actively seek out published content for removal" (Klonick, 2018, p. 1635). In their detection efforts, hosts may rely on the content that users upload as well as the metadata associated with that content, such as account information, IP address, volume/frequency of posting, and other signals. Efforts to detect "coordinated inauthentic behavior" on social media services, for example, have primarily relied on the use of metadata (François, 2020).

During the **evaluation** phase, the user-generated content is examined to determine whether it does violate the host's policies, or is potentially a violation of a relevant law. Evaluation can be done by humans, automatically, or through a combination of automated and human review. For example, an image that has been reported by a user may be examined by a human moderator or run through a hash-matching program (see p. 22 below) to determine whether the image matches content the provider already knows it wants to block. When hosts employ automated content filtering to block content at upload, the separate steps of detection, evaluation and enforcement may collapse into a single step.

**Enforcement** is the action a moderator takes against user-generated content that it determines violates a content policy or law. While removing content is one possible enforcement method, there are a wide variety of other actions moderators can take against violative content (Goldman, 2021). For example, moderators can: add a warning before users may access the content or counterspeech such as a fact-check; disable user comments or other features for a post; decrease the availability of some or all of a user's posts, such as through shadowbans, downgrading content's visibility in search results, or restricting forwarding or sharing of posts; impose monetary remedies, such as demonetizing content; or suspend or deactivate a user's account (Goldman, 2021; Masnick, 2018).

## E. Appeal

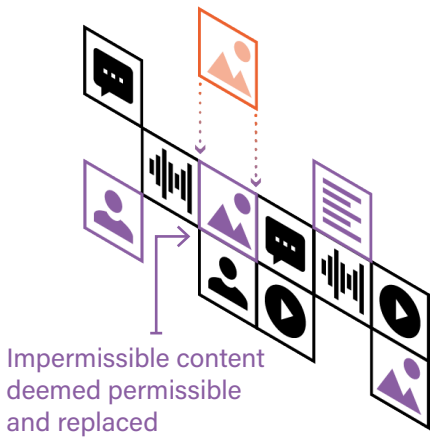

<span style="color:purple">Impermissible content deemed permissible and replaced</span>

## F. Education

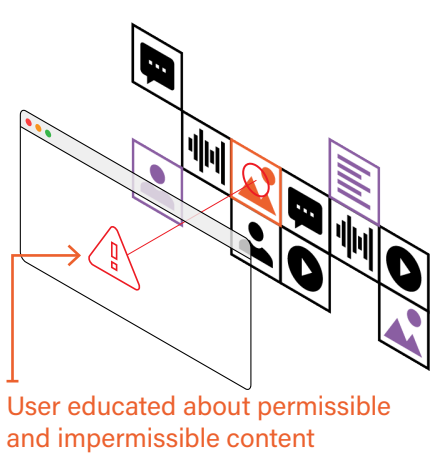

<span style="color:orange">User educated about permissible and impermissible content</span>

After enforcement, some hosts allow users to **appeal** or otherwise seek review of content moderation decisions that users believe are erroneous (Klonick, 2020). Errors are inevitable in content moderation, and given the amount of content some hosts moderate, even those with a very high rate of accuracy in content moderation will still make thousands or millions of erroneous decisions.[3] Accordingly, appeals are an essential part of content moderation. In addition to allowing user appeals, content moderation policies should recognize the inevitability of errors and build in opportunities to review and refine both content moderation policies themselves and the tools used to implement them.

Finally, hosts can **educate** users about their content moderation policies and the ways in which the policies are enforced. The education phase can take various forms. The most basic are the service's terms of service, Community Guidelines, and other user-facing information about the site's policies. Moderators may also educate users on permissible and impermissible content by "praising good behavior and criticizing bad" or otherwise providing users with explanations when their content is moderated (Grimmelmann, 2015, pp. 61–63). This can include the notifications sent to users when they are informed that action has been taken against their content, or when an appeal they have made has been denied. Education is a critical component of content moderation, especially in response to good faith violations of content moderation policies. Users must understand what kinds of content are and are not allowed for the content moderation process to function effectively (Jhaver et al., 2019).

The preceding discussion on the interrelated phases of content moderation are relevant for all types of settings and relied on examples from plaintext (or non-encrypted) environments in particular. As we noted earlier, our goal here is to understand the implications of encryption on these phases. To do that, the next section explains what we mean by encryption before moving into our analysis of proposals that seek to enable some forms of content moderation in end-to-end encrypted environments.

---

3   For example, Klonick (2020) explains that in the second and third quarters of 2019, less than 1% of the 4.8 billion pieces of content that Facebook removed from the site were appealed, resulting in the restoration of 10.1 million pieces of content (pg. 2433-34).

# Understanding End-to-End Encryption (E2EE)

Encryption is crucial for protecting end-user privacy and security, as well as the promotion of free expression online (Kaye, 2015). However, challenges remain in ensuring that end-users understand exactly what encryption is, including specifically end-to-end encryption (Bai et al., 2020). Policy-makers likewise encounter many myths about end-to-end encryption (Global Encryption Coalition, 2020).

We start our analysis by first defining encryption with reference to its key cryptographic characteristics: signing and encryption (Uhlig et al., 2021). An encryption scheme takes as input a key and a message and outputs a ciphertext (an encrypted version of the message). Encryption schemes vary and may offer different properties. For example, an encryption scheme can use a key to package the data in a way that ensures authentication (knowing who sent it), confidentiality (only the receiver can open it) and integrity (message was not tampered with). Given the key, the ciphertext can be decrypted to reveal the message. Encryption is rarely used in isolation and is usually one part of a larger system like a messaging app or cloud storage service. In such systems, encryption is used to guarantee the authenticity, confidentiality and integrity of data.

How exactly encryption is implemented in the system depends on who gets access to the data and how. Because encryption can only guarantee confidentiality between parties that share an encryption key, one of the most important considerations when deploying encryption is how to manage key sharing. This shared exchange, coupled with the confidentiality of encryption, allows system designers to protect data in transit, while it is at rest, and between end points.

Encryption in transit means that encryption is deployed to protect data while it is being communicated over a network by the service provider. More precisely, it means that the data's transport is encrypted between nodes authenticated by the communication channel and hidden from all other nodes that might be relaying the data. This is the case, for example, when a web browser encrypts a user's web traffic before sending it to a web server. Here, the decryption keys are known only to the web server.

**A system, service, or app is end-to-end encrypted if the keys used to encrypt and decrypt data are known only to the sender and the authorized recipients of this data.**

Encryption at rest refers to the use of symmetric encryption to protect data while it is stored. More specifically, it means that a user employs one key that they may or may not share with others so that the data is decrypted when they want to use it and encrypted when it is not in use. As an example, consider a cloud storage service that backs up and stores user data. In such a service, the data is sent from the user device to the cloud provider—usually with encryption in transit—and then encrypted by the provider to be stored. If the provider's data centers are breached (virtually or physically), access to the storage data would be denied without the symmetric encryption key that only the user has. In such a deployment, another user or the cloud provider could be authorized to gain access to the data by being given knowledge of the key, but ideally they are not. While encryption in transit and encryption at rest are essential to building a secure system, they protect data in these specific contexts. To secure the data that is exchanged between two or more users who are not intermediary service providers, and such that intermediary service providers cannot witness the conversation, another architecture is required.

That architecture, for deploying encryption that protects user data at all times and against any on-path attacker, is end-to-end encryption (E2EE). Here, data is encrypted on the user's device and can only be decrypted by authorized users who have exchanged keys with one another. Because these users are the only ones with knowledge of the decryption keys, data in E2EE systems is confidential to those users and no one else, not even the intermediary service provider (Knodel et al., 2021).

In summary, a system, service, or app is end-to-end encrypted if the keys used to encrypt and decrypt data are known only to the sender and the authorized recipients of this data. Specifically, this implies that intermediate parties that route, store, backup, and process the encrypted data do not have access to the keys and, therefore, cannot learn any information about the data.

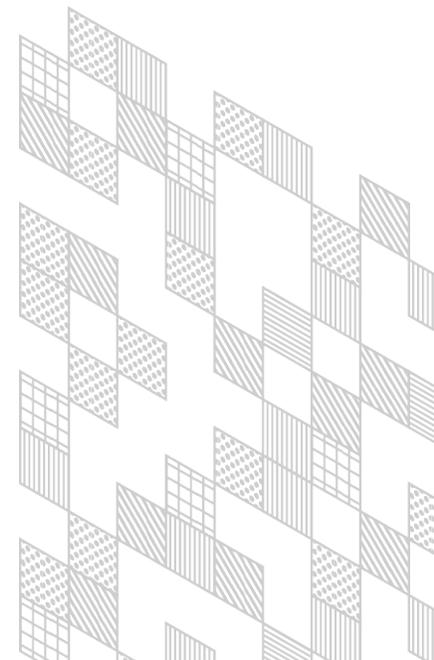

## Examples of Services that include E2EE

In practice, end to end encryption is used in a number of ways, all of which have attracted the concern of law enforcement and others in terms of content detection. Below are descriptions of a few common services and the privacy guarantees that E2EE is meant to provide for them:

*Storage*. This is a cloud storage service that stores end-to-end encrypted files or photos. That is, the data is first encrypted under a key known only to the user and then stored in the cloud. Examples include services like the Keybase file system and the Pixek photo app. In this setting, E2EE is used to guarantee that the user is the only party that can access the data. Other examples like Dropbox encrypt at-rest user files and photos but the service retains access to the key.

*Messaging*. An encrypted message exchange is a conversation between two or more people over an end-to-end encrypted messaging app. Here, messages are encrypted using keys known only to the participants in the conversation. This includes messaging apps like WhatsApp and Signal. Here, E2EE is used to make the conversation confidential in the sense that only the authenticated participants in the conversation can access the messages.

*Email*. This is an email service, or tool, that allows users to send and receive end-to-end encrypted emails. Like encrypted messaging, the keys are only known to the sender and recipients, which guarantees that no third party—even the email server—can see the contents of the email.

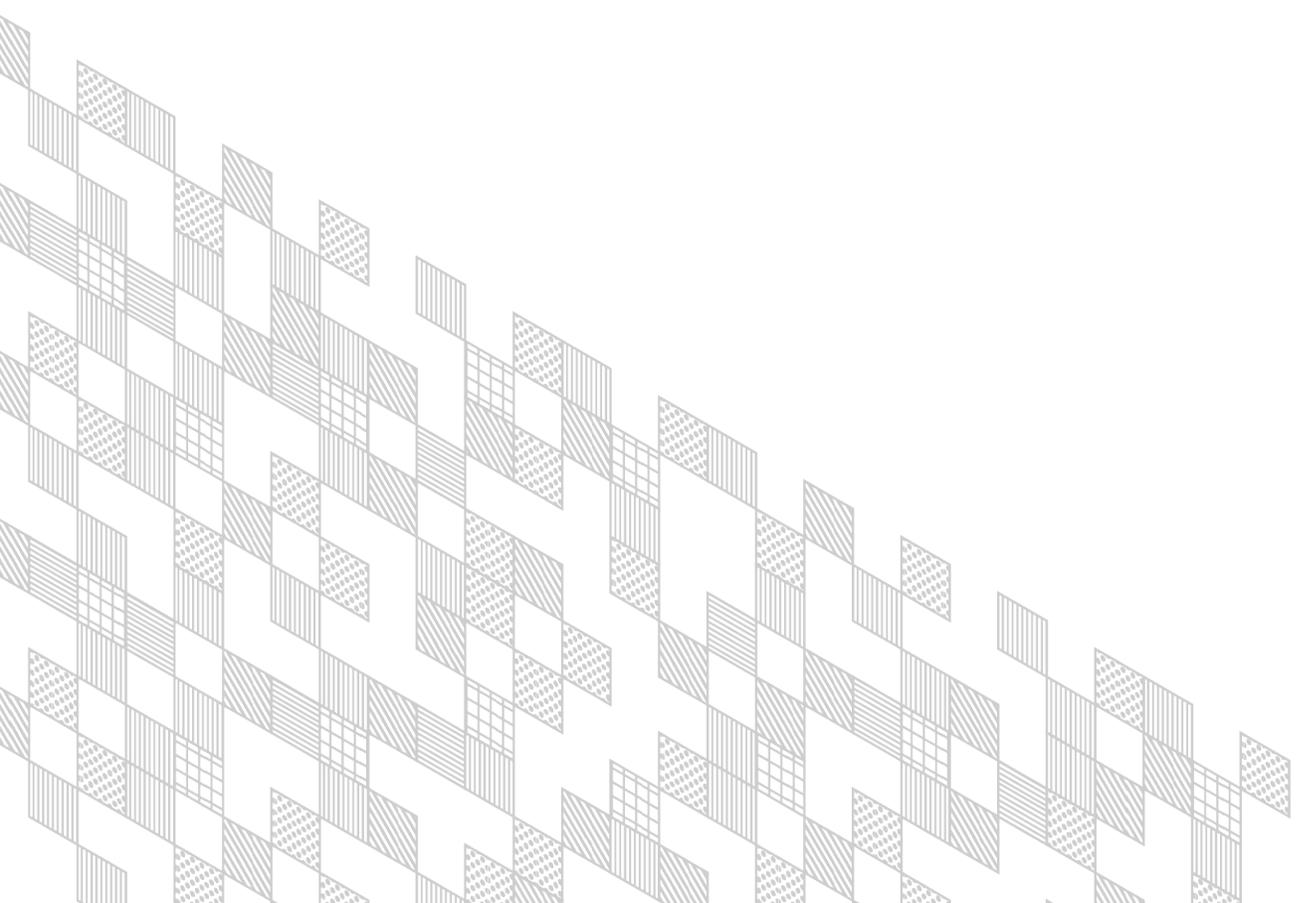

# Detecting Content in E2EE Environments

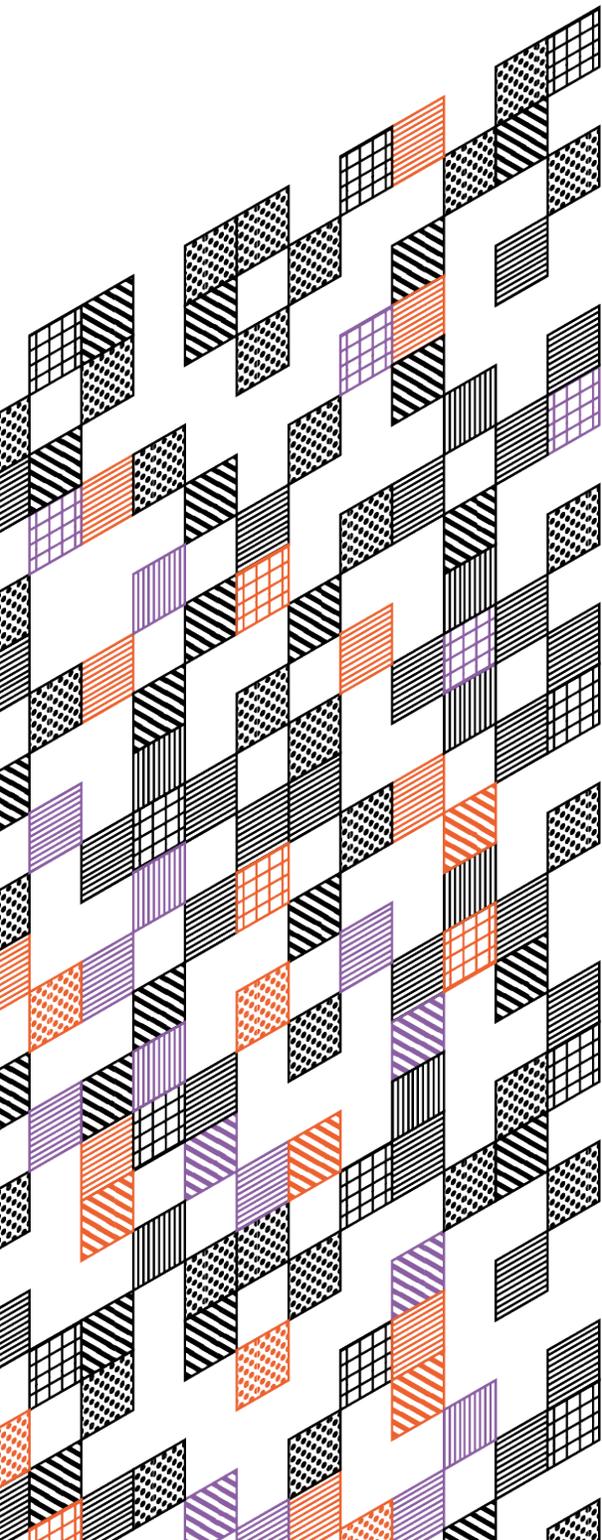

With the expanding array of end-to-end encrypted services available, law enforcement concerns about "going dark" have returned to the forefront of policy debates around the world.[4] A new angle on these deliberations is the focus on content moderation, raising questions of whether and how providers of E2EE services can or should proactively identify problematic content, respond to abusive uses of their systems, and implement legal orders to block content.

As policymakers engage in these important debates about how, for example, to stop the spread of child sexual abuse material or to address abuse of E2EE services by terrorist organizations, it is important they understand that questions of how best to deal with problematic content in E2EE environments are often specific to how the encryption is implemented. It is also essential to center the privacy and security guarantees system designers and users expect from end-to-end encryption, namely that only the sender and authorized recipients have access to the encryption and decryption keys and the data, and therefore intermediate parties do not. With this in mind, we review a number of technical proposals emerging from research in computer science and cryptography that seek to enable content detection in E2EE services.[5]

One initial observation about these proposals is that they are focused on content detection: how can a service provider identify that some subset of encrypted data is problematic content?

---

4    In a plaintext environment, law enforcement or other government agencies often gain access to the content of private communications to investigate crimes and protect national security, with or without the cooperation of the service provider. With E2EE, this is typically not possible and so some governments have called for measures that would give law enforcement and security agencies "exceptional access" to E2EE communications (National Academies of Sciences, Engineering, and Medicine, 2018). Such "exceptional access" proposals would give law enforcement access to the keys used to decrypt the data using a key escrow system, or enable encryption "backdoors" by deliberately modifying the encryption scheme to allow third-party access (e.g., by the service provider in response to a legal process). More specifically, where service providers enable such forms of exceptional access this means that, by definition, we are no longer talking about an E2EE system. It also means that other actors (and potential adversaries) besides law enforcement and security agencies may gain access to the content (Abelson et al., 2015). In general, introducing backdoors will intentionally introduce vulnerabilities into a system, increasing risks for all users (Global Encryption Coalition, 2020).

5    See Appendix for further technical details of select proposals.

For most types of abusive, illegal, or otherwise "harmful" content, there can also be significant disagreement across countries and communities about how to define such prohibited content.

**What is often framed as a debate about moderation of unwanted content in E2EE services is really a discussion about (any) content detection in E2EE.**

As discussed earlier, detection is only one phase of content moderation. Furthermore, the types of content of interest—typically "harmful," illegal, or otherwise unwanted content such as terrorist propaganda, CSAM, mis- and disinformation, or spam—have no technically unique characteristics that make them readily distinguishable from more innocuous types of content (i.e., an image is an image regardless of its content). Thus, what is often framed as a debate about moderation of unwanted content in E2EE services is really a discussion about (any) content detection in E2EE.

Below, we examine five types of techniques used in both E2EE and plaintext/unencrypted systems, which generally attempt to detect content already uploaded or added to the system, and/or attempt to restrict unwanted content from being added to the system. These are user reporting, traceability, meta-data analysis, perceptual hashing, and predictive models.

# User Reporting

Within a plaintext environment, several options already exist for detecting unwanted content that has already been uploaded or posted to the host service. One of the most common approaches is to enable some form of user-initiated reporting. A service provider may make tools available (e.g, reporting buttons, complaint forms, contact information) that allow users to alert moderators, other intermediaries, or other users of unwanted content. Moderators are able to directly view content, and either take action or escalate it for further review. Law enforcement agencies or other third-parties may establish dedicated reporting channels with service providers, in order to notify providers of potentially illegal content.[6] And providers may become aware of unwanted content on their service through emails, news reports, or other communications that occur outside of the specific flagging tools/procedures that the provider has developed.[7]

Using these options as a starting point, researchers and others have put forward several proposals to enable user-reporting in E2EE services. There are currently a variety of proposed cryptographic schemes to enable the reporting (by users, to service providers) of end-to-end encrypted messages. These solutions are designed so that the messages can only be decrypted and verified by the service provider and no one else beyond the original sender and recipients. This category of schemes is called "message franking." Given a private conversation between users A and B, message franking guarantees that:

1. B can prove to the service provider that they received a given message from A; and

2. B cannot claim to the service provider that they received a message from A that they never received.

Message franking is already in use in E2EE systems. For example, Facebook employs message franking for their end-to-end encrypted messaging system, Secret Conversations (Facebook, 2016). After Facebook introduced message franking, follow-up work improved upon their original scheme by making it more efficient for file attachments (Dodis et al., 2018; Grubbs et al., 2017), allowing only partial opening of messages (only specific pieces of a message are revealed) (Chen & Tang, 2018; Leontiadis & Vaudenay, 2018), and extending message franking to metadata private service providers, i.e., service providers that do not reveal who the sender and recipients of messages are (Tyagi, Grubbs, et al., 2019).

---

[6] Note that this is controversial.

[7] Note that this is a separate question from whether a service has actual knowledge that a specific post is illegal.

Although message franking enables the service provider to view messages, this does not violate the properties we expect of an encrypted conversation since one of the original participants in the conversation (here, the receiver) explicitly chooses to reveal the message to the service provider. From a more technical perspective, the keys used for encryption and decryption are still only held by the sender and the receiver, and there are no backdoors that enable the host or any other third party to access the conversation without the knowledge and approval of at least one of the participants. All that message franking permits is for one party in a communication to disclose it to the service provider in a way that the service provider can be sure of the message's authenticity.

**Message franking explicitly binds senders to their messages so that they can be held accountable if they send unwanted or illegal content such as hate and harassing messages, CSAM, terrorist propaganda, and/or spam.**

With message franking, a private conversation is no longer deniable to someone that was part of that conversation (i.e., the designated moderator can verify a reported message). This is useful where non-repudiation, or the need to ensure that the author of a message cannot deny their authorship, is important. In fact, message franking explicitly binds senders to their messages so that they can be held accountable if they send unwanted or illegal content such as hate and harassing messages, CSAM, terrorist propaganda, and/or spam. This technique could also potentially allow the person reporting the message to protect themselves from liability where they receive abusive or illegal content; for example, message franking allows the person reporting the message to verify that they were the recipient, not the originator, of a message containing CSAM.

While message franking can improve the functionality of user-reporting for problematic content in E2EE, not all E2EE services have this feature. For example, WhatsApp and Signal both use the Signal protocol where deniability is actually viewed as an important property to promote privacy (Marlinspike, 2013). In a system built with deniability in mind, where A sends a message to B, B can authenticate that the message is from A, but cannot prove this to anyone else.

Although the message franking techniques described above do not violate the end-to-end guarantee of a private conversation, newer variants could, so it is important that practical deployments of message franking be transparent about the exact properties they guarantee.

# Traceability

A related concern is how to identify which users have shared content that has been flagged as problematic. For example, the service provider or a government agency may want to trace how a particular piece of content was distributed online. Because standard end-to-end encryption makes detection and tracing harder, researchers are studying the extent to which tracing can be done over end-to-end encrypted messaging platforms.

One proposal by Tyagi, Miers, et al., (2019), extends the techniques from message franking to trace all the users who forwarded or received a given piece of content. Prior to someone flagging the content, the messages are kept confidential and only the sender and recipient(s) of a message are able to decrypt it. After the content is reported, however, the service provider can learn the contents of a conversation and trace it to find all the messages containing the same content that were not directly reported in the forwarding chain. Although this allows a service provider to trace the spread of viral malicious content, it also provides an opportunity for users to report sensitive content and expose the privacy of all senders and recipients in the chain.

These tracing techniques are built on top of message franking and, while franking does not violate the properties we expect from encrypted conversations, tracing does. Recall that in the baseline use of message franking, the message can only be revealed to the service provider if one of the participants in the conversation chooses to report it, thus preserving the privacy guarantee of E2EE that only the sender and authorized recipients have access to the data.

With tracing, however, the service provider can learn information that was not explicitly revealed to it by either the sender or receiver. For example, if user A sends a message to user B who then sends the same message to user C, user C can report the message and the tracing scheme will reveal not only that B sent the message to C but that it was sent by A to B without either A or B explicitly revealing that to the service provider. In a more expanded forwarding chain, an individual who may be 1000 links removed from the original sender may report a message, which would then reveal that the message was received by the previous 999 recipients.

The demand for traceability among governments is often based on proposals that are politically motivated with little, if any, technical guidance in terms of feasibility. Last year, the Brazilian government introduced its so-called "Fake News Law" requiring a form of traceability and user identification which would not only break encryption and undermine privacy and free speech, but also place a significant burden on service providers to retain large amounts of user data (Maheshwari, 2020).

In another example, the Indian government recently enacted "The Information Technology (Intermediary Guidelines and Digital Media Ethics Code) Rules, 2021" that undermines free expression (Maheshwari & Llansó, 2021), while requiring large social networking services, those with more than 5 million registered users, to reveal the original sender or originator of a given message.

The government advocated two potential approaches to doing this: requiring providers to include the originator's encrypted identity information in the metadata of each message; or requiring providers to maintain a hash database of all content conveyed on their platform. When problematic content is identified, its hash can be matched against the hashes in the database to identify the originator. These proposals, as well the rules themselves, are flawed in several ways because the concept of an originator is ambiguous, the proposals are not practically feasible, and the rules are legally questionable (Maheshwari & Nojeim, 2021). Most importantly, while the proposal may avoid identifying everyone in a chain of messages (unlike the Tyagi, Miers, et al., (2019) proposal), it does reveal the identity of the originator to a third party, undermining the privacy guarantee of E2EE.

**Traceability as a concept is not consistent with the privacy guarantees for E2EE systems and fixing design issues in these flawed examples won't resolve this inherent tension.**

In sum, depending on the approach, traceability can provide a third party with access to information about the originator of a message and/or all other prior recipients of the message in an expanded forwarding chain without their consent. Either way, this means that traceability as a concept is not consistent with the privacy guarantees for E2EE systems and that fixing design issues in these flawed examples won't resolve this inherent tension.

# Metadata Analysis

Analysis of metadata, or "data about data" which in this case is data about an encrypted message, can include a surprisingly robust amount of detail including file size, type, date/time, sender/receiver, etc. Analysis based on metadata is relevant, for example, in the detection of spam in plaintext communications. With E2EE, a service provider could also detect spam (or similar problematic content) by examining the volume or size of messages sent by an account and take action if this volume deviates from the provider's classification of normal messaging activity.

Machine learning (ML) techniques could be applied to metadata to predict the extent to which a given user may share problematic content on an E2EE service. ML is a process by which a system parses data to extract characteristics and relationships within the data. For example, analysis of unencrypted profile, or group chat information (e.g., profile pics) using ML classifiers can contribute to the detection of accounts involved in the distribution of CSAM on E2EE services. According to WhatsApp, they ban more than 300,000 accounts per month for suspected sharing of CSAM using these kinds of approaches (WhatsApp, 2021b).

These techniques could focus on user behavior as well. For example, ML models can be trained on the behavior of users that have been banned from an E2EE service (e.g., their account creation practices, frequency of sending messages, or reports from other users about problematic content). This can then be used to analyze the behaviour of new users who wish to join the service or existing users. (WhatsApp, 2019)

**In general, as long as the metadata analysis occurs exclusively on a user's device and does not store, use, or send decrypted messages, the user's privacy is preserved and the guarantees of end-to-end encryption are not violated.**

Of course, not all metadata analysis may reliably identify problematic content. The WhatsApp client on a user's device places a label on content that is forwarded (shared) with users many times (WhatsApp, 2021a). However, its utility as a content detection technique is limited because messages that are frequently forwarded are not necessarily problematic—they might simply be interesting.

Moreover, it is also important to recognize the privacy risks inherent with the access of metadata by service providers as it can be used to reveal sensitive data such as the identities of the sender or receiver (Greschbach et al., 2012). These risks are present in the use of metadata analysis in plaintext communications and potentially in E2EE systems as well. Limiting metadata analysis to the user's device (and to an app on that device) can be one approach to reducing those risks. In general, as long as the metadata analysis occurs exclusively on a user's device and does not store, use, or send decrypted messages, the user's privacy is preserved and the guarantees of end-to-end encryption are not violated.

# Perceptual Hashing in E2EE

There are two broad categories of machine learning tools used for content detection and analysis: matching models and predictive models (Shenkman et al., 2021), both of which have been proposed for E2EE systems. ML matching models aim to recognize content as identical or sufficiently similar to content it has seen before. A key technique is hashing, an approach for creating a digital fingerprint or representation of a piece of content for purposes of comparison in a way that is more efficient and flexible than relying on the original content for comparison. These digital fingerprints can be compared with each other to identify matches. To be useful in detecting unwanted content at scale, content hosts rely on databases of hashes of previously identified problematic content. The content host runs the hashing algorithm on every user-provided file at upload and compares that hash to the hashes in the database.

There are two main types of matching models: cryptographic hashing and perceptual hashing. Cryptographic hashing uses a cryptographic function to generate a random hash fingerprint, which is highly sensitive to change. This approach can be effective in identifying known content without alterations. Perceptual hashing, on the other hand, allows the service provider to determine the degree to which two pieces of content must be similar in order to be deemed a match. This can be important where minor changes are made to a piece of content to bypass detection.

Perceptual hashing is used in a plaintext context to automatically identify content that the host has previously determined it does not want on its system; see for example Cloudflare's CSAM scanning tool, (Paine & Graham-Cumming, 2019). It is also the type of hashing that is the subject of research proposals for content detection in E2EE. Perceptual hashing can be used by a service provider when it receives content; this is called server-side scanning. Or it can be used by the messaging app (or another app) on the user's device before the content is ever sent; this is called client-side scanning. In either case, if a match is found, the message can be blocked from reaching the recipient and the user may face additional consequences.

In the case of encrypted conversations, one could use either server-side or client-side scanning to detect unwanted or abusive content, but these approaches either violate the privacy guarantees one expects from an encrypted conversation, introduce new security vulnerabilities, or both. In the case of server-side scanning, the content would be hashed before being encrypted and the hashes would be sent to a platform server to be checked. Revealing the hash to the server, however, is a privacy violation because hashes can reveal information about the content. For example, someone with access to the server could create hashes of specific images or other content of interest, and when there is a match they could then determine who sent that content to the server.

There are a few hash databases that are shared across the industry, including NCMEC's database of alleged CSAM images and the Global Internet Forum to Counter Terrorism's shared industry database of potential terrorist and violent extremist content. Beyond those shared resources, some services may create their own hash sets of content they wish to block, for example, Facebook's program for hashing user-submitted intimate images that are being nonconsensually shared on the service. https://about.fb.com/news/2019/03/detecting-non-consensual-intimate-images/.

**In the case of encrypted conversations, one could use either server-side or client-side scanning to detect unwanted or abusive content, but these approaches either violate the privacy guarantees one expects from an encrypted conversation, introduce new security vulnerabilities, or both.**

To address these limitations, one proposal suggests a privacy preserving perceptual hashing approach which would allow a platform server to check whether some encrypted content matches known unwanted material without learning anything else about the material itself (Kulshrestha & Mayer, 2021). Furthermore, since this solution is server-side, the database of unwanted hashes would be kept only by the platform and not distributed to users' devices. While such solutions try to minimize the amount of information revealed to the server, learning that an encrypted message matches or does not match unwanted content is a violation of the privacy we expect from encrypted storage and conversations, as a third party now has access to some information about the communications. These solutions also can produce false matches, which, depending on the purpose of the hash database, may mean that the service provider will interpret the matched content to be abusive; therefore, it is crucial that they include ways to redress false matches.[8]

Conversely, with client-side scanning, the set of unwanted hashes are stored on the user's device so that the hash comparisons can be done on the device. If the results of the hash comparison are only provided to the user, this may not violate the privacy guarantees we expect from an encrypted conversation; however, if the results of the hash comparison are shared with the server, then the privacy guarantees of E2EE have been violated. There are several additional concerns about perceptual hashing in plaintext that raise questions about its efficacy. For example, it is only effective on content that is shared more than once. One study found, however, that 84% of CSAM images that were reported (either by US service providers using automated detection tools, or by the U.S. public) were only reported once, demonstrating that much of the abusive material is new and, therefore, not something that could be blocked by a matching tool (Bursztein et al., 2019).

In addition, hash filtering, particularly where the algorithm is public, is also vulnerable to the deliberate addition of hashes to the database to generate false positives, i.e., a poisoning attack (Dolhansky & Ferrer, 2020). Poisoning attacks can be used to censor speech by adding hashes of politically sensitive material into a hash database. Researchers have already found some evidence of this in China where actors may be using client-side scanning for this purpose (Knockel et al., 2020). And even if client- and server-side scanning are used only to detect whatever may be defined as abusive or harmful content in one jurisdiction, building the technology into platforms and user devices could allow this kind of exploitation by authoritarian (or non-authoritarian) governments elsewhere (Pfefferkorn, 2020).

**While such solutions try to minimize the amount of information revealed to the server, learning that an encrypted message matches or does not match unwanted content is a violation of the privacy we expect from encrypted storage and conversations, as a third party now has access to some information about the communications.**

---

8    The solutions try to keep the false match rate of perceptual hashing over end-to-end encrypted data about the same as perceptual hashing on plaintext data.

The practical implementation of client side scanning in E2EE introduces these vulnerabilities into the system. In particular, distributing the database of unwanted hashes to user devices could allow bad actors to subvert the detection process by manipulating the hash database.

Another proposal by Reis et al., (2020) explores the idea of using perceptual hashing for detecting misinformation on WhatsApp, and, by extension, other E2EE services, using a client-side scanning model. Their work mainly involved understanding information sharing patterns on WhatsApp, but in the process proposed an architecture that could be introduced in WhatsApp to detect and flag misinformation on user devices. In their proposal, Facebook would maintain a set of perceptual hashes for images that fact-checkers have deemed to be misinformation (e.g., images shared out of context or manipulated using simple techniques to create so-called "cheap-fakes" (Paris & Donovan, 2019)). These hashes are then stored directly on a user's device and periodically updated. Upon sending an image, its hash would be compared to the ones already stored on the sender's device and further warnings or notifications can then be displayed to the sender if the content has been identified as misinformation. The same check can be done on the recipient's device and with similar warnings and notifications displayed.

Under this proposal, users sending this information would not be flagged by the platform. The notification actions are only taken on the sender and recipients' devices. Recipients may then choose to report the message but there is no automated means of accountability for sending or receiving misinformation under this proposal. However, while this approach to client-side scanning may protect user privacy (detection is only done on the user's device and no third party is involved), it is also vulnerable to circumvention, which could limit its utility overall. In a client-side scanning implementation, the full set of hashes to be blocked is stored on each user's device, making that hash set potentially discoverable to a malicious user. It may be possible for a malicious adversary to use this set of hashes to identify what images are reflected in the database (for example, by hashing images they want to share and determining whether there is a matching hash in the database). This could allow the malicious adversary to develop methods of applying transformations to content that are reflected in the database, in order to avoid detection.

**In a client-side scanning implementation, the full set of hashes to be blocked is stored on each user's device, making that hash set potentially discoverable to a malicious user.**

This weakness in the client-side scanning model would apply to most forms of client-side scanning. It is particularly problematic in use cases involving the sharing of illegal media such as CSAM, where malicious actors may be highly motivated to develop circumvention methods. In fact, this partly explains why algorithms for well known methods such as PhotoDNA[9] are not public or run locally on devices as they may be vulnerable to attack.

In addition to these limitations, there may be other deployment considerations such as the device's processing power, storage, internet connectivity, and battery usage. These may have important equity implications depending on the context where this approach is used. For example, among low-income populations, countries, or regions, feature phones or low-end smartphones are often used with WhatsApp rather than more powerful smartphones like iPhones and Samsung devices that are popular in high-income countries (James, 2020).

In sum, hash-matching techniques such as server-side scanning or client-side scanning either provide a third party with access to the message, introduce significant security vulnerabilities into the system, or both. Even proposals for client-side scanning that involve no third-party access (i.e., only the sender and/or receiver are notified about the detection of unwanted content) introduce the potential for manipulation of the hash database by bad actors. These approaches are therefore not consistent with the privacy and security guarantees of E2EE.

---

9    A tool developed by Microsoft and Dartmouth College to detect CSAM using perceptual hashing (Microsoft, n.d.).

# Predictive Models for Content Detection in E2EE

The second category of ML techniques consists of prediction models that aim to recognize the characteristics of content based on the machine's prior learning. This approach is often used for content that is new or previously unknown. It requires (often large amounts of) data to train the model to predict whether a piece of content has certain attributes. These include computer vision models, which cover the analysis of shapes, textures, colors, etc., and computer audition models, which focus on audio content. A basic example of one such technique, an image classifier, may seek to predict whether an image uploaded by a user is a dog or cat. In a plaintext setting, ML tools are used for detection and identification of text-based content (see Duarte et al., 2017) and multimedia content (Shenkman et al., 2021).

Building on these techniques, Mayer (2019) offers guidelines for researchers to develop models to predict the existence of problematic content in an E2EE context. One way to do this could use machine learning algorithms to detect plaintext problematic messages (like spam) by using pre-trained classifiers installed, for example, on a messaging or other app on a user's device. Once a user decrypts a message, the classifier can flag the message as potential spam. Again, there are practical limitations to consider such as the phone's battery life and processing capabilities. As with metadata analysis, if this process occurs exclusively on a user's device and no information about the message is disclosed to a third party, then the guarantees of end-to-end encryption may not be violated. However, more research is needed to develop viable techniques using this approach.

# Content Moderation in E2EE Environments - Next Steps for Research

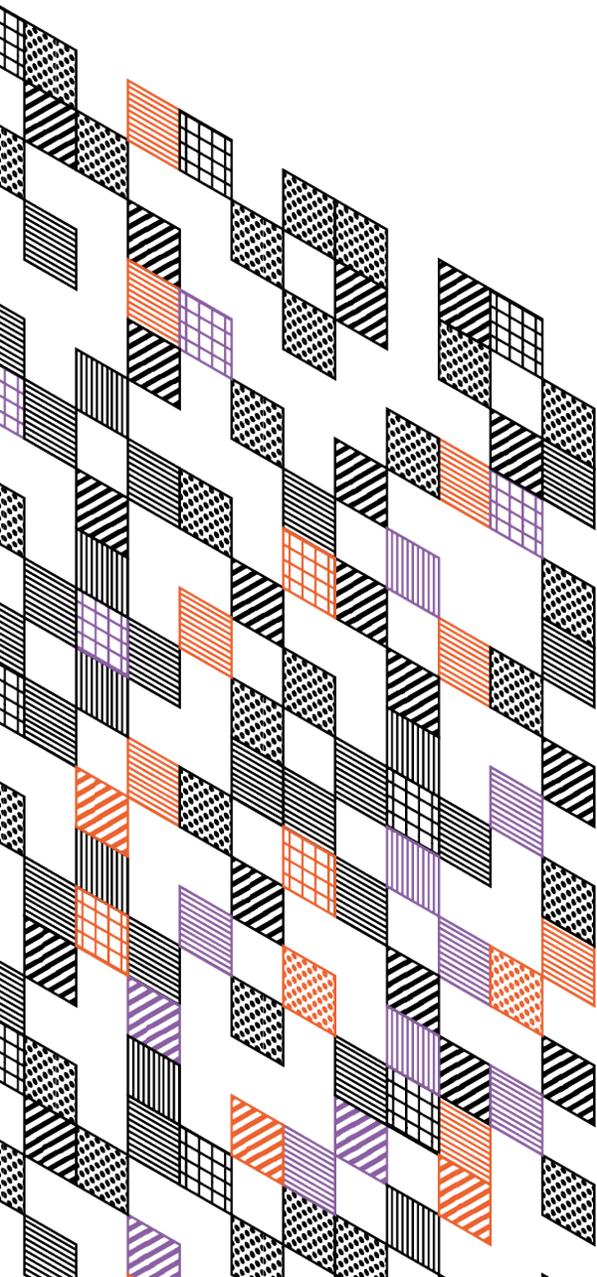

**W**e defined E2EE environments as a service or app where the keys used to encrypt and decrypt data are known only to the senders and designated recipients of this data. A crucial part of this is the end-to-end principle. Third parties that route, store, backup, and process the encrypted data do not have access to the keys and, therefore, cannot learn any information about the data.

Using this definition, we assessed current technical proposals that aim to provide some form of content detection in E2EE services. Our assessment identified two content detection proposals that preserve the security and privacy guarantees of E2EE without introducing any new security vulnerabilities into the system. The first is user reporting, which includes message franking, a means for the service provider to authenticate that the sender actually sent content that was reported as problematic by the receiver. Message franking enables user reporting of problematic content such as abusive content, mis- and disinformation, or CSAM, including in encrypted one-to-one and group chat settings. The second approach to content detection that is consistent with the promises of E2EE is the use of metadata analysis, which could be used, for example, to detect problematic content such as spam and CSAM.

Several other proposals for enabling content detection in end-to-end systems exist, but for different reasons, they introduce new vulnerabilities into the system, are unable to provide the privacy and security guarantees the user expects, or are not yet viable. While some may appear promising at first (e.g., Reis et al., 2020), they introduce vulnerabilities such as those presented by having the hash database on the user's device. Others effectively break the privacy guarantees of E2EE (e.g., traceability). And still other potentially promising approaches (Mayer, 2019) such as ML classifiers that operate solely on the user's device and disclose no information to third parties are not yet available and point to areas for investigation particularly in terms of predictive models for detecting abusive content on the device.

In general, we believe that there are significant opportunities for researchers to improve content detection and other aspects of content moderation while fully respecting the guarantees of E2EE services. As a next step, we suggest that researchers consider these broad guidelines:

- Explore the applicability of content moderation approaches beyond just content detection in E2EE services. Every E2EE service is different, but some tools and approaches that are useful in the plaintext context could be relevant in end-to-end encryption environments. User education about a service's policies or a particular forum's rules, options for users to act as moderators in multi-person discussions, and accurate evaluation of reported content are all potential intervention points in a content moderation system. Automated tools that are currently configured to detect, evaluate, and enforce against content in a single process can be reconfigured and incorporated into moderation systems that rely more heavily on user reports. These types of interventions may be especially useful in thinking about how to design E2EE services to reduce the likelihood of abusive content and activity, such as harassment and hate speech.

- Content detection solutions should emphasize user agency, which is the case with user reporting, including message franking. Metadata analysis combined with user-reporting can allow the user to determine appropriate actions where problematic content is detected. This could also include further research on allowing users to choose or create their own filters to block unwanted content within the E2EE app on their device, as long as these filters do not expose information about the messages to third parties.

- If E2EE services will need to rely substantially upon user reports to detect unwanted or potentially illegal content, then significantly more research is needed to determine the most effective techniques for encouraging user reporting of content such as user interface design, alignment of users with the service's values, and promotion of the development of healthy communities.

- Additional research is needed to prevent abuse by repeat offenders including users that have been banned from creating new accounts by an E2EE or similar service.

- Proposals must be explicit about the exact properties they guarantee, and that any change to a system needs user notification, consent and opt-out. It's essential to base research on the end-to-end principle that it is possible to maintain the security and privacy guarantees that the user expects.

**User-reporting and metadata analysis provide effective tools in detecting significant amounts and different types of problematic content on E2EE services including abusive and harassing messages, spam, mis- and disinformation, and CSAM.**

In sum, we note that user-reporting and metadata analysis provide effective tools in detecting significant amounts and different types of problematic content on E2EE services including abusive and harassing messages, spam, mis- and disinformation, and CSAM. However, more work is still needed. We encourage additional research to be done on content moderation in E2EE services based on the use of metadata itself and methods that act within the confines of a messaging app on a user's device to empower the user to flag, hide, or otherwise report unwanted content to the service provider. These methods should neither modify the underlying encryption schemes in any way, nor encroach on the privacy and security guarantees of end-to-end encryption. Ultimately, we should recognize that technological solutions to detecting problematic content alone, whether in a plaintext or E2EE system, will not address the larger issues of, say, the distribution of disinformation or CSAM. Rather, as a society, we also need to consider the social and political causes behind these phenomena and address them at their core.

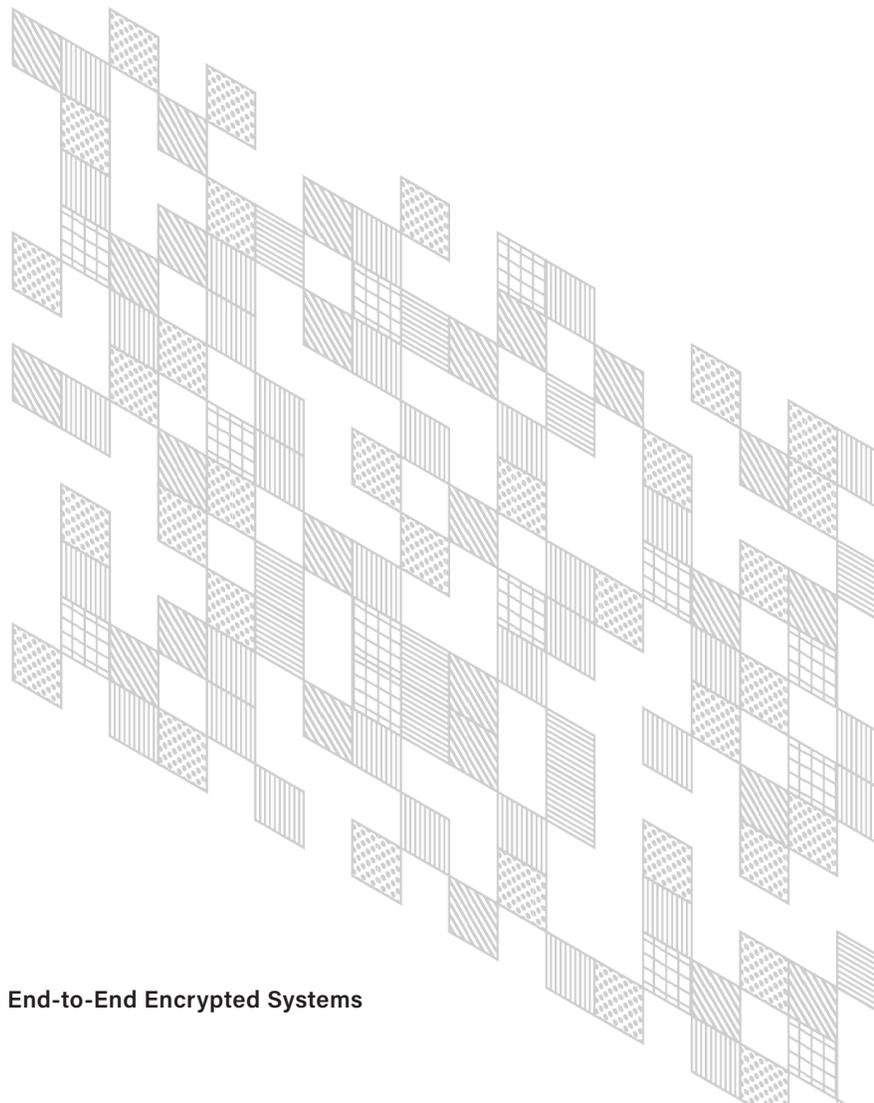

# Appendix: Extended Summaries of Select Proposals to Detect Content on E2EE platforms

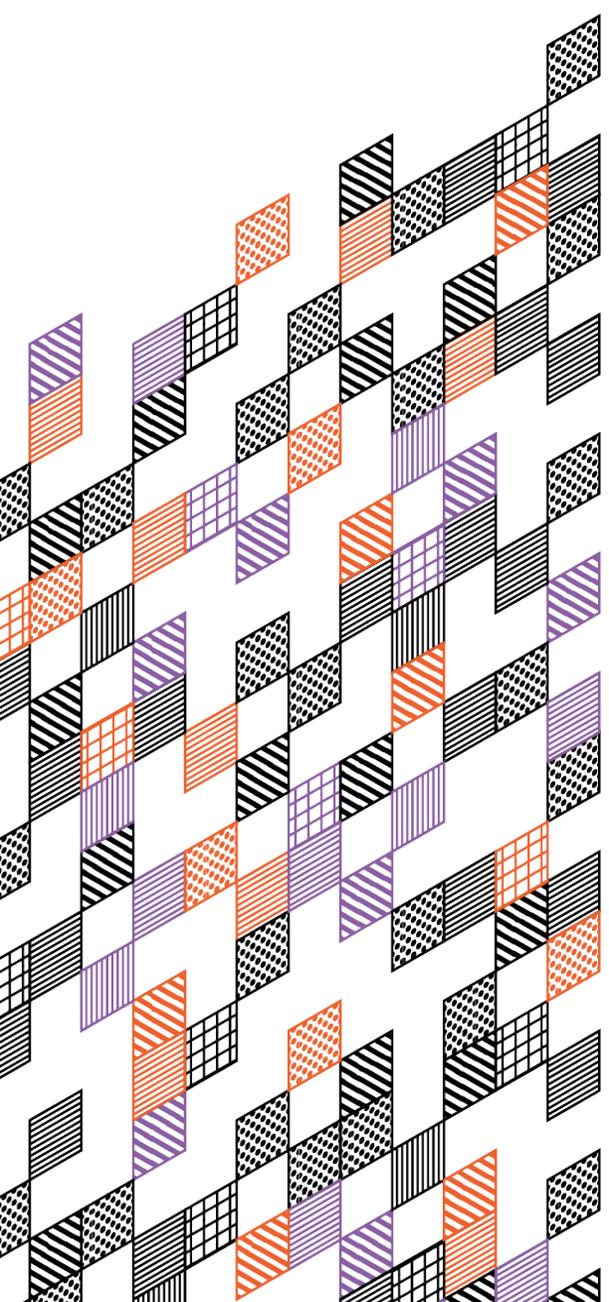

### Grubbs, P., Lu, J., & Ristenpart, T. (2017)

The authors formalize a security definition for message franking and evaluate Facebook's scheme. In the process, they also formalize the properties of sender binding, which ensures that once a message has been committed to, a commitment will later verify correctly given the original message, this prevents a sender from being able to send a message to the recipient that properly decrypts but fails to verify when reported; and receiving binding, which ensures that once a recipient receives a commitment for a message, they should not be able to open that commitment to any other message than the one originally sent, this prevents the receiver from being able to frame the sender for content that they did not send.

This builds upon the previous work by introducing a more efficient scheme that allows for the franking of file attachments. Though the previous scheme may be efficient for text messages, large file attachments containing images or videos require a faster franking mechanism. The authors are directly motivated by Facebook's file attachment franking scheme, which although is efficient, does not fulfill the previously discussed sender binding property.

### Dodis, Y., Grubbs, P., Ristenpart, T., & Woodage, J. (2018)

The authors first demonstrate an attack against Facebook's attachment franking scheme. They do so by exploiting the fact that Facebook uses AES-GCM, a secure AEAD scheme, in a non-standard setting. A malicious message sender under Facebook franking scheme is able to construct ciphertexts in a manner that prevents a recipient from reporting a malicious message, thus violating sender binding security. This attack is carried out by a malicious sender first creating two different messages, m1,m2, where m1 is innocuous and m2 contains malicious content. The malicious sender can then find two keys, k1,k2, such that the encryption of both messages, under different keys, results in the same ciphertext (Enc(k1,m1) = Enc(k2,m2)). The sender sends the innocuous message first and then the malicious message. When encrypted, these two messages produce identical ciphertext, therefore Facebook may internally assume the second message is a duplicate. If Facebook delivers the second message but does not tag it, a later abuse report from the recipient would fail.

**Dodis, Y., Grubbs, P., Ristenpart, T., & Woodage, J. (2018) (Continued)**

The authors note that this demonstrates the need for a more efficient means of committing file attachments without violating sender binding security, which motivated them to create a one-time secure primitive called encryptment. They then construct a scheme called hash function chaining that allows for the franking of a file attachment with a single SHA-256 or SHA-3 computation. The underlying security comes from a property of the types of hash functions used, called collision resistance. Under collision resistance, a computationally-bounded malicious adversary cannot find two inputs that hash to the same output. By leveraging this property, the authors chain together multiple hash function calls, with the xor of a key and attachment as the input. This then produces a binding commitment to a given file attachment.

Using this scheme, they are able to achieve unforgeability, sender binding security, receiver binding security, and deniability. They also leverage the intermediary outputs of computing a commitment as key material for encrypting the attachment itself.

**Chen, L., & Tang, Q. (2018)**

In prior works franking necessitates that a recipient of malicious content reveal the full message contents when reporting a message. This may lead recipients to refrain from reporting a message containing malicious content in fear of revealing other sensitive information. A malicious message sender can also take advantage of this and send malicious messages with sensitive information about the recipient in order to deter the recipient from later reporting these messages. To address these issues, the authors create a scheme that allows for partial messages to be revealed when reporting.

**Tyagi, N., Grubbs, P., Len, J., Miers, I., & Ristenpart, T. (2019)**

Asymmetric message franking extends prior work in this area to cover messaging platforms that are metadata private. In the previous context, if user A sends an end-to-end encrypted message to user B, the platform cannot uncover the contents of the message unless a user reports the message. The platform does however see that a message was sent from user A to B. In the context of metadata privacy, when user A sent a message to B, the platform only sees that a message was sent, and not the identities of the sender or receiver. The previous proposals use the user's identity as part of the commitment to a message and therefore would not work in this setting.

This work introduces an asymmetric message franking scheme that is compatible with metadata private end-to-end encryption messaging platforms that fulfills unforgeability, receiver binding, sender binding, and deniability. With metadata private end-to-end encryption platforms, users each have two keys, a private key and a public key. If user A were to send user B a message, A first must digitally sign the following and create a commitment using A's private key: the message, B's public key, the moderator's public key. A moderator can later verify this commitment using their private key, the public keys of A and B and the reported message.

**Kulshrestha, A., &
Mayer, J. (2021)**

This work shows that perceptual hashing can still occur on top of end-to-end encrypted messages in a manner where a server only learns whether the content within an encrypted message matches against known harmful content without learning what the user's original message or the hash. The user also does not learn anything about the contents within the database. The authors make use of a cryptographic technique called private information retrieval that allows for an element to be retrieved from the database without the database knowing what the element was.

Though this mitigates the risk of revealing hashes to the server, the same risks in enabling surveillance still exist. Even a protocol that does not reveal any information about the original message can be abused by platforms to conduct surveillance on top of end-to-end encryption. Instead of finding matches to harmful content, a platform could also substitute in other content that may be unfavorable toward the platform or perhaps a particular government. The authors also note that the deployment of these protocols in the context of a democratic government may enable more authoritarian governments to use the same tools in order to suppress speech and conduct surveillance.

**Tyagi, N., Miers, I., &
Ristenpart, T. (2019)**

The authors begin by introducing a scheme that allows for the backward tracing of malicious content but since this does not capture all of the users who may have received the content, they then introduce a second scheme. A moderation system can then trace backward and forward all the individuals who may have received the malicious content and notify users as well as identify the source of the content. Prior to a user report, confidentiality is preserved and only the sender and recipient of a message are able to decrypt it. After a report is made, the platform, through tracing, learns the content of messages that were not directly reported in the forwarding chain. Although this allows a platform to trace the spread of viral malicious content, it also provides an opportunity for malicious users to report sensitive content and expose the privacy of all senders and recipients in the chain.

The authors achieve the ability to trace backward and forward malicious content by introducing two separate schemes for (1) backward tracing and then (2) both backward and forward tracing. In the first scheme, the authors create a chain of what they call encrypted pointers in which forwarded messages point back to the previous sender of the message. When a message is sent, the sender samples a tracing key and uses it to create a commitment to the message. If a user A were to send a message (m) to user B who then forwards the (m) to user C, user A first randomly samples a tracing key kA and uses it to create a binding commitment to (m). This commitment, commA, all

**Tyagi, N., Miers, I., & Ristenpart, T. (2019) (Continued)**

along with kA are sent to B. When B forwards (m) to C, B samples its own key, kB and then encrypts kA using kB. This creates what is called an encrypted pointer, which points back to the previous sender. When initiating a backwards trace from C, kB is used to retrieve kA, which can be used to verify A's commitment to (m). This scheme works for forward tracing using the same concept and creates encrypted pointers that point to recipients as well as senders.

The authors are able to achieve receiver binding and sender binding in a similar manner to previous work by relying on the collision resistance property underlying their commitment scheme. They also maintain deniability by only allowing the platform to perform tracing.

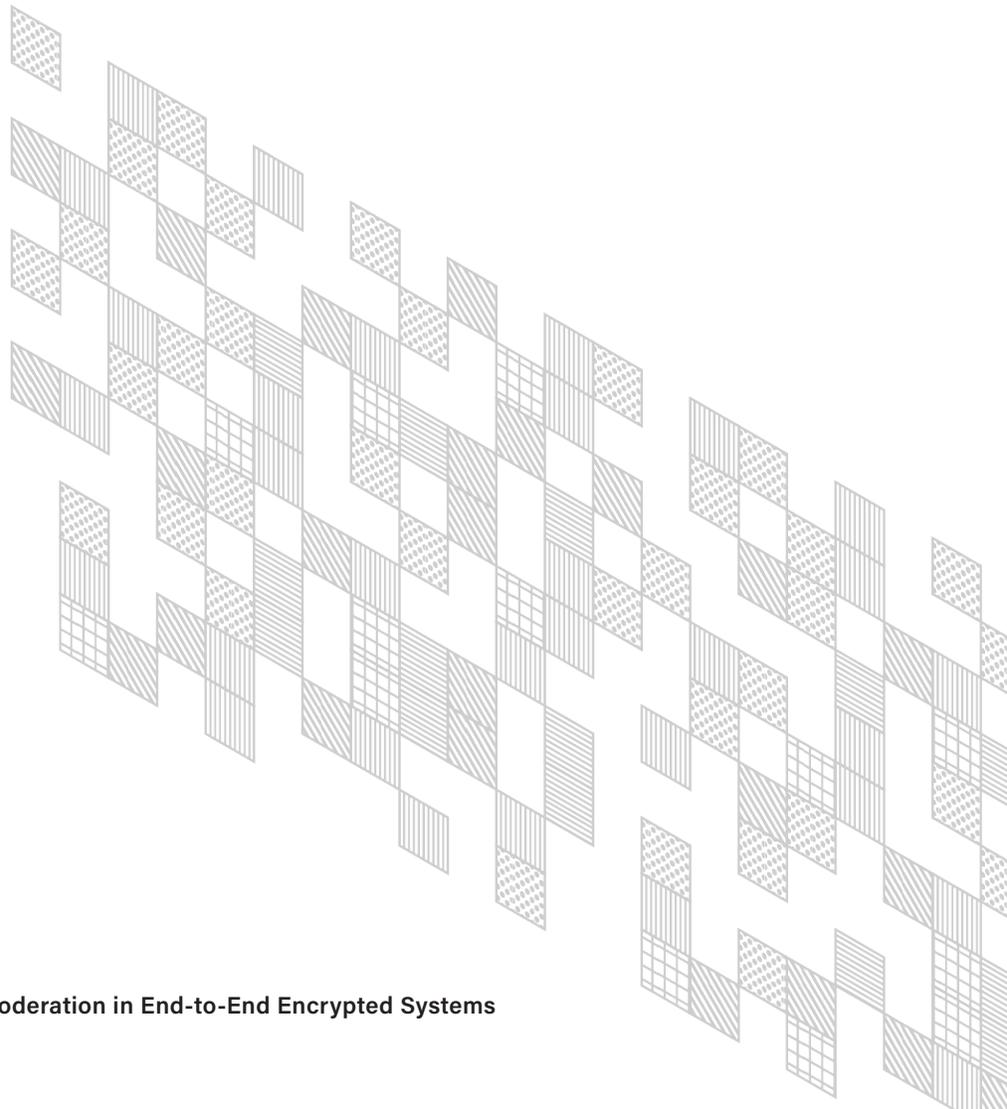

cdt.org

cdt.org/contact

Center for Democracy & Technology
1401 K Street NW, Suite 200
Washington, D.C. 20005

202-637-9800

@CenDemTech